\documentclass[runningheads]{llncs}
\usepackage[T1]{fontenc}
\usepackage{graphicx}
\usepackage{amsmath,amsfonts}
\usepackage{algorithmic}
\usepackage{textcomp}
\usepackage{xcolor}
\usepackage{hyperref}
\usepackage{listings}

\lstdefinelanguage{json}{
    basicstyle=\ttfamily\footnotesize,
    numbers=none,
    stepnumber=1,
    numbersep=5pt,
    showstringspaces=false,
    breaklines=true,
    frame=single,
    rulecolor=\color{black},
    backgroundcolor=\color{gray!10},
    stringstyle=\color{red},
    keywordstyle=\color{blue}\bfseries,
    commentstyle=\color{gray},
    morestring=[b]",
    morestring=[d]',
    morecomment=[l]{//},
    morecomment=[s]{/*}{*/},
    literate=
     *{0}{{{\color{orange}0}}}{1}
      {1}{{{\color{orange}1}}}{1}
      {2}{{{\color{orange}2}}}{1}
      {3}{{{\color{orange}3}}}{1}
      {4}{{{\color{orange}4}}}{1}
      {5}{{{\color{orange}5}}}{1}
      {6}{{{\color{orange}6}}}{1}
      {7}{{{\color{orange}7}}}{1}
      {8}{{{\color{orange}8}}}{1}
      {9}{{{\color{orange}9}}}{1}
      {:}{{{\color{blue}:}}}{1}
      {,}{{{\color{blue},}}}{1},
}

\begin{document}
\title{Malware Detection based on API calls}

\author{Christofer Fellicious\inst{1}\orcidID{0000-0001-7487-7110} \and
Manuel Bischof\inst{2}\and
Kevin Mayer \inst{2}\and
Dorian Eikenberg \inst{2}\and
Stefan Hausotte\inst{2}\and
Hans P. Reiser\inst{1,3}\orcidID{0000-0002-2815-5747} \and
Michael Granitzer \inst{1} \orcidID{0000-0003-3566-5507}
}
\authorrunning{C. Fellicious et al.}
\institute{University of Passau, Germany \email{firstname.lastname@uni-passau.de}\and
GData Cyberdefense, Germany
\email{firstname.lastname@gdata.de}\\ \and
Reykjavik University, Iceland\\
\email{hansr@ru.is}}

\maketitle              % typeset the header of the contribution
\newcommand{\gdata}{G DATA CyberDefense AG}
\begin{abstract}
Malware attacks pose a significant threat in today's interconnected digital landscape, causing billions of dollars in damages. Detecting and identifying families as early as possible provides an edge in protecting against such malware.
We explore a lightweight, order-invariant approach to detecting and mitigating malware threats: analyzing API calls without regard to their sequence.
We publish a public dataset of over three hundred thousand samples and their function call parameters for this task, annotated with labels indicating benign or malicious activity.
The complete dataset is above 550GB uncompressed in size.
We leverage machine learning algorithms, such as random forests, and conduct behavioral analysis by examining patterns and anomalies in API call sequences.
By investigating how the function calls occur regardless of their order, we can identify discriminating features that can help us identify malware early on.
The models we've developed are not only effective but also efficient. They are lightweight and can run on any machine with minimal performance overhead, while still achieving an impressive F1-Score of over 85\%.
We also empirically show that we only need a subset of the function call sequence, specifically calls to the ntdll.dll library, to identify malware.
Our research demonstrates the efficacy of this approach through empirical evaluations, underscoring its accuracy and scalability. 
The code is open source and available at Github\footnote{\url{https://github.com/cfellicious/api-based-malware-detection}} along with the dataset\footnote{\url{https://zenodo.org/records/11079764}}.
%Our goal is to set a new standard for API-based malware detection, thereby significantly enhancing security within API-driven environments.
\keywords{machine learning \and dataset \and API based malware detection.}
\end{abstract}
\section{Introduction}
\label{sec:introduction}
Malware attacks are increasing all over the world year over year\cite{av_test_statistics}. 
Zero-day exploits, a term used to describe vulnerabilities in software that are unknown to the software vendor and therefore unpatched, also help malicious attackers hide their presence on hijacked machines. 
These exploits pose a significant challenge for malware detection as they can be used to launch attacks that are difficult to detect and defend against. 
%With polymorphic malware, which changes its code multiple times to evade detection, the task of detecting malware is becoming more complex. 
Polymorphic malware presents a significant challenge for traditional code-based checks, as the code itself gets encrypted, making it difficult to identify and analyze.
With the introduction of polymorphic malware into the wild, regular code-based checks do not work, as the code itself gets encrypted, along with garbage values inserted into the malware payload files.

Machine learning shows promise in identifying malware but requires large amounts of data to generalize well. 
Although many cybersecurity companies use machine-learning-based methods, most datasets are proprietary. 
Existing datasets are often proprietary, narrow in scope, or insufficiently comprehensive, posing challenges for researchers and developers aiming to create and refine machine learning models for malware detection. 
While some publicly available datasets exist, they tend to be outdated, small in scale, or lacking in diversity, limiting their utility in addressing modern malware's sophisticated and rapidly evolving nature. 
This highlights a critical need for the development and dissemination of large-scale, diverse datasets that encompass a wide range of benign and malicious software behaviors. 
We address this gap by creating a dataset from data collected with the help of~\gdata~and making it publicly available.

Another aspect is the practical side of malware detection. 
System software running on machines should have a manageable overhead and not impede the performance of essential business software.
Therefore, having very complex models for real-time analysis that take up valuable resources is not feasible, as the memory and performance requirements will hinder other software.
We could have more complex offline models for forensic analysis. However, the best option on a running machine would be a lightweight model that uses simple feature engineering but offers a very high degree of accuracy.

Our primary contribution is the creation of the largest publicly available dataset of API calls, with over 300,000 malware samples and 10,000 benign samples of API call instances sourced from recent malware and benign software samples. 
The uncompressed size of the dataset exceeds 550GB and is available on Zenodo.
The second contribution is lightweight models that could be used in real-time malware detection based on their API calls.
We empirically show that we could detect malware reliably with as few as 250 API calls.
Our findings show that the mere frequency of API calls is a powerful indicator of malicious intent, enabling the detection of malware with high certainty.
Moreover, our research underscores the importance of comprehensive data collection in developing robust malware detection systems, providing a valuable resource for the cybersecurity community.
For our third contribution, we conduct a comprehensive study to determine the number of API calls required for the best detection performance at different maximum API call counts.

We introduce the premise and research gap in section~\autoref{sec:introduction}.
We discuss the current landscape of malware detection methods in~\autoref{sec:related_work}.
Our dataset creation and malware identification method is present in~\autoref{sec:method}. We present the results of our experiments in~\autoref{sec:results}~and our conclusions in~\autoref{sec:conclusion}.
\section{Related Work}
\label{sec:related_work}
Aboaoja et al. published a survey on the various issues and challenges of Malware Detection~\cite{aboaoja2022malware}. The authors outline that detecting evasive malware is still one of the biggest challenges. Although there are several approaches to detecting evasive malware, such as using multiple execution environments to identify evasive behavior, the time and resource complexity lead to each approach having its weakness.
Shukla et al. developed a method that uses RNN to detect the so-called stealthy malware~\cite{shukla2019stealthy}. The authors define stealthy malware as "malware created by embedding the malware into a benign application through advanced obfuscation strategies to thwart the detection." The authors "translate the application binaries into images, further convert it into sequences, and extract local features for stealthy malware detection."
Feng et al. proposed the method DawnGNN using Graph Attention Networks~(GAT)~\cite{feng2024dawngnn}. The authors proposed a novel documentation-augmented Windows Malware Detection Framework. The method works by converting the API sequences into API graphs to extract contextual information. The authors encode the functionality descriptions using BERT and finally use Graph attention for classification.
Li et al. proposed a method to detect dynamic malware based on API calls~\cite{li2022novel}. The authors used \textit{intrinsic features} of the API sequence. The authors claim that this allows the models to capture and combine more meaningful features. 
The authors then use the category, action, and operation object of the API to represent the semantic information of each API call. 
The authors do the classification using a Bidirectional LSTM module and their results outperform the baselines.
Cui et al. proposed a graph-based approach to detect malware from API-based call sequences~\cite{cui2023api2vec}. 
The proposed method works by creating two graphs, a Temporal Process Graph~(TPG)~and a Temporal API Graph~(TAG)~to model intra-process behavior. 
A heuristic random walk algorithm then generates several paths that can capture the malware behavior.
The authors generate the embeddings using the paths pre-trained by the Doc2Vec model.
Chen et al. proposed a parameter-augmented approach for the Microsoft Windows platform called~\textit{CruParamer}~\cite{chen2022cruparamer}. The method employs rule-based and clustering-based classification to compute the sensitivity of an API-call parameter to malicious behavior.
The classification is done by concatenating the API embedding to the sensitive embedding of the labeled APIs so that their relationship is captured in the input vector.
The authors then train a binary classifier to identify malware, and according to the authors, their model outperforms naive models.
Almousa et al. proposed a method to identify ransomware attacks based on API calls~\cite {almousa2021api}. The authors initially studied the lifecycle of ransomware on the Microsoft Windows platform. The next step was to extract malicious code patterns. The authors used data from publicly available repositories and sampled the malicious code in a sandbox. Machine learning models were built based on this analysis and yielded a high detection rate.

There are also multiple public malware datasets based on API calls.
Catak et al. published a malware dataset that contained different malware types~\cite{catak2019benchmark}. The dataset contains eight malware types, namely, Trojan, Backdoor, Downloader, Worms, Spyware Adware, Dropper, and Virus, for 7107 samples. 
The authors created the dataset using the Cuckoo Sandbox, available on GitHub.
Zhang published a dataset of 10,654 samples with sample labels~\cite{zhang22}.
The author divides the dataset into normal, ransomware, miner, DDoS Trojan, worm, infective virus, backdoor, and Trojan.
This dataset is from the Alibaba Security Algorithm Challenge.
Trinh created a much larger dataset of 1.55 million samples, of which 800,000 malware and 750,000 "goodware" samples are present~\cite{trinh21}.
Another dataset is from Oliveira with "42,797 malware API call sequences and 1,079 goodware API call sequences"~\cite{de2023behavioral}.
The dataset comprises API call sequences of the "first 100 non-repeated consecutive API calls associated with the parent process, extracted from the 'calls' elements of Cuckoo Sandbox reports".
A few other open datasets from challenges, such as the Aliyun Malware Detection Dataset~\cite{tianchi2016} exist, but most of the datasets are either unavailable publicly or have a too narrow scope.
More recent work is by Maniriho et al., where the authors created their dataset~\cite{maniriho2023api}. The dataset consists of 1285 malicious and 1285 benign samples. The dataset is publicly available and hosted on GitHub.
\section{Method}
\label{sec:method}
We structure this section into two parts. The first describes the dataset, its creation, and corresponding statistics. 
The second explains the order-invariant method we use to generate the results.
\subsection{Dataset}
\label{subsec:dataset}
%With the dataset, we aim to address a few gaps in the existing landscape of malware-based API call datasets. 
%\todo{This is the direction we want to go}
%We wanted to create a dataset that, 
A popular method to detect malware includes tracing the Application Program Interface~(API)~calls.
An API, or application programming interface, is a set of rules or protocols that enables software applications to communicate with each other to exchange data, features, and functionality~\footnote{\url{https://www.ibm.com/topics/api}}.
We created this dataset to address the gaps in the existing landscape of the current malware-based API call datasets with the following properties,
\begin{itemize}
    \item up to date with the current malware
    \item large and varied enough to cover most modern malware
    \item does not restrict the labels to a single category. Malware might not always fall into a specific category alone. From a cybersecurity perspective, it makes more sense to group samples by the malware family.
    \item a dataset collected from the real-world machines. We worked closely with~\gdata~for data collection. 
    Working directly with a cybersecurity firm gives us the advantage of knowing the malware is from the real world.
\end{itemize}

%A popular method to detect malware includes tracing the Application Program Interface~(API)~calls.
%An API, or application programming interface, is a set of rules or protocols that enables software applications to communicate with each other to exchange data, features, and functionality~\footnote{\url{https://www.ibm.com/topics/api}}.
In a Microsoft Windows environment, user processes interact with the Operating System using dynamically linked libraries~(DLL). 
The Windows~\emph{ntdll.dll}~library is one such library.
"The NTDLL runtime consists of hundreds of functions that allow native applications to perform file I/O, interact with device drivers, and perform interprocess communications"~\cite{ms_ntdll}. 
Any malware or, for that matter, the user process will need to communicate with the NTDLL library, which makes it the perfect library for hooking our API call logger.
This library contains hundreds of functions related to different functionalities, such as semaphores, threads, creating events or objects, and so forth.
The library belongs to the Native API and can call functions in user or kernel mode~\cite{chappell2024}.
Therefore, we decided to trace the function calls to the ntdll.dll library.
Tracing all the functions to the ntdll.dll is cumbersome as documentation only exists for a few functions.
Therefore, we select a subset of~$59$ proven valuable functions of the NTDLL library~\cite{inside_ntdll}.
We obtained this set of function calls in cooperation with the cybersecurity experts from~\gdata.%~\cite{gdata}.

We wanted to keep up with the latest malware trends, so we collected samples for approximately six months between 01.05.2023 and 01.11.2023.
%The meticulous collection of these samples will provide us with a comprehensive understanding of the trends and ensure we have a diverse range of malware samples.
During this period, we undertook the monumental task of observing approximately $1,000,000$ malicious samples.
%We further truncated the list of these samples to around~\textbf{330,000}~samples, as we did not want to oversaturate the dataset by a few classes. 
Malware occurs in waves in the real world. If a specific malware type is successful, it spreads like wildfire across different networks.
This, in turn, caused a spike in a few classes in our dataset, and we removed such samples to avoid biasing any classifier to a few classes.
We collected these samples~(approximately 330k), and we grouped the samples into different malware families.
%These malware samples were then collected and split into families.
Labeling this data proved to be a challenge as well.
However, the most common form of labeling data for malware is grouping it into predefined classes such as trojan, virus, backdoor, rootkit, and so on.
But newer malware could have overlapping patterns with multiple labels and having a single label might not capture the complete behavior of the malware.
Therefore, rather than group them into a single category, such as a virus, backdoor, or trojan, we group them into distinct malware families based on source code analysis from~\gdata.
This means that malware belonging to the same label in our dataset will exhibit the same properties and this should allow for better grouping and analysis by cybersecurity researchers.

For a malware detection dataset, we also require data from benign software as well.
Acquiring benign software is not a problem at all, as there are millions of verified benign software from different organizations all over the world.
Our problem was that most benign software required user interaction to install or had a Graphical User Interface~(GUI).
The drawback of having such benign software in the dataset was that delineating between malware and benign software would be easy as most malware does not have a GUI.
We needed benign software that did not have a GUI and required no user interaction to execute.
This criterion alone made it similar to some types of malware as some malware genres also require no input from the user~\cite{gopinath2023comprehensive,bilot2024survey}.
%Our cybersecurity partner firm maintains a whitelist of benign software which we used in addition to Microsoft service executables to generate the API traces for the benign samples.
Our cybersecurity partner firm maintains a comprehensive whitelist of benign software, which includes a variety of trusted applications and services regularly used in enterprise environments. This whitelist was instrumental in ensuring the integrity and reliability of our benign sample set. In addition to this whitelist, we included Microsoft service executables, which are widely recognized as baseline components of the Windows operating system. We aimed to cover a broad spectrum of typical, non-malicious software behavior by incorporating these executables.
%To generate the API traces for our benign samples, we executed each application and service in a controlled environment, meticulously recording all API calls made to the ntdll.dll library. This process involved monitoring the software in real-time to capture an accurate representation of its interaction with the operating system. By doing so, we ensured that our dataset reflected genuine benign activity, providing a robust foundation for distinguishing between normal and malicious behavior.
Including a diverse set of benign software, sourced from our partner's whitelist and Microsoft services, significantly enhances the robustness of our dataset. This diversity is not just a factor, but a key element for training machine learning models that need to accurately differentiate between benign and malicious API call patterns. Our approach ensures that the benign samples are representative of real-world software environments, thereby improving the reliability and effectiveness of the malware detection system we developed.

%Each malicious and benign sample in our dataset has a unique SHA value, which identifies the executable.
%Malware belonging to the same family can have different executable versions, determined by their SHA values, based on various factors, such as different versions, code rewrites, or slight changes to defeat detections via SHA analysis.
%This differentiation of executables based on their SHA values is important as it allows us to keep track of mutating malware samples.
%We execute each malicious sample in a virtual environment and monitor the process.
%The tools used for monitoring and logging are proprietary to our cybersecurity partner.
We uniquely identify each malicious and benign sample in our dataset using a SHA value, which serves as a digital fingerprint for each executable.
This SHA value is crucial for distinguishing between different versions and variants of software, particularly for malware samples that belong to the same family but exhibit variations in their code. 
These differences can arise from factors such as version updates, code rewrites, or intentional modifications designed to evade detection mechanisms that rely on SHA analysis.
By assigning a unique SHA value to each executable, we can accurately track and manage the diversity of samples within our dataset. 
This differentiation is significant for mutating malware, which frequently changes its code to avoid detection. 
The ability to identify and catalog these variations ensures that our dataset reflects the dynamic nature of real-world malware, enhancing the robustness of our detection models.
We utilize a controlled virtual environment to execute and monitor each malicious sample. 
This approach allows us to observe the malware's behavior in isolation, preventing any potential damage to actual systems. 
During execution, we meticulously monitor and log all API calls made by the malware to the ntdll.dll library. 
This process's monitoring and logging tools are proprietary to our cybersecurity partner, ensuring precise and secure data collection.
These proprietary tools are specifically designed to capture detailed API call traces, providing a comprehensive view of the malware's interaction with the operating system. 
We then use the resulting data to construct a detailed profile of each sample's behavior, which is integral to developing our machine learning-based malware detection system. 
By leveraging unique SHA values and advanced monitoring techniques, we ensure that our dataset is extensive and precise, forming a solid foundation for accurate and reliable malware detection.

We simulate an internet connection using an open source library~\footnote{\url{https://github.com/prskr/inetmock}}.
Some malware samples, after unpacking themselves, request the download of an executable. We simulate the executable download by sending a predefined harmless executable whose functionalities are inert and whose behavior is predictable.
We also monitor any direct child, determined by~\textbf{InheritedFromUniqueProcessId}, from the~\textbf{EPROCESS}~struct~\cite{eprocess}.
Therefore, if the malicious sample forks, we will still monitor all the child processes and log their API calls.
Each such sample, both malware and benign, is traced over a runtime of $360$ seconds.
The execution environment is a Microsoft Windows 10 21H2 virtual machine with 4GB RAM and one vCPU.
The logger then dumps all the calls and call parameters into a JSON file named SHA.
We show the sample trace of a single API call in~\autoref{fig:json_log}.
\iffalse
\begin{figure*}[tb]
    \centering
\begin{minted}[frame=single, rulecolor=black, linenos=false, style=murphy,
label=Sample call format]{json}
{
    "level":"info",
    "ts":"2023-12-04T09:23:36Z",
    "msg":"Monitored function called",
    "vmi_ts":"2023-12-04T09:19:09Z",
    "vmi_logger":"ApiTracing_FunctionHook",
    "vmi_Parameterlist":"
    [
        {\"SystemInformationClass\":192},
        {\"SystemInformation\":1373520},
        {\"SystemInformationLength\":32},
        {\"ReturnLength\":0}
    ]",
    "vmi_FunctionName":"NtQuerySystemInformation", 
    "vmi_ModuleName":"ntdll.dll",
    "vmi_ProcessDtb":"65aca001",
    "vmi_ProcessTeb":"2e6000"
}
\end{minted}
\caption{Sample log of a single API Call}
%\Description[a short description]{Sample log of a single API Call}
\label{fig:json_log}
\end{figure*}
\fi

\begin{figure*}[tb]
    \centering
\lstset{
  language=JSON,
  frame=single,
  rulecolor=\color{black},
  numbers=none,
  basicstyle=\ttfamily\footnotesize,
  backgroundcolor=\color{gray!10},
  captionpos=b,
  breaklines=true,
  showstringspaces=false,
  keywordstyle=\color{blue}\bfseries,
  stringstyle=\color{red},
  commentstyle=\color{gray},
}

\begin{lstlisting}[label=Sample call format]
{
    "level":"info",
    "ts":"2023-12-04T09:23:36Z",
    "msg":"Monitored function called",
    "vmi_ts":"2023-12-04T09:19:09Z",
    "vmi_logger":"ApiTracing_FunctionHook",
    "vmi_Parameterlist":
    [
        {"SystemInformationClass":192},
        {"SystemInformation":1373520},
        {"SystemInformationLength":32},
        {"ReturnLength":0}
    ],
    "vmi_FunctionName":"NtQuerySystemInformation", 
    "vmi_ModuleName":"ntdll.dll",
    "vmi_ProcessDtb":"65aca001",
    "vmi_ProcessTeb":"2e6000"
}
\end{lstlisting}

\caption{Sample log of a single API Call}
\label{fig:json_log}
\end{figure*}

We created an online repository containing all information about the dataset and our code along with the names of the traced functions\footnote{\url{https://github.com/cfellicious/api-based-malware-detection}}.
Overall, the uncompressed size of the dataset is approximately 572GB in total. We already published the dataset on Zenodo\footnote{\url{https://zenodo.org/records/11079764}}, and the dataset is currently public access.

\iffalse
\begin{figure}
    \centering
    \includegraphics[width = 0.99\linewidth]{figures/API_calls_per_file.png}
    \caption{Distribution of API Calls per malware sample}
    \label{fig:api_calls_per_file}
    %\Description[a short description]{lorem ipsum dolor}
\end{figure}

We created an online repository containing all information about the dataset and our code along with the names of the traced functions\footnote{\url{https://github.com/cfellicious/api-based-malware-detection}}.
\fi
\subsection{Order invariant Method}
We aim to develop a malware detection method independent of the temporal constraints and ordering of API calls. This approach ensures that the detection system remains effective even when the sequence of API calls is altered, which is a common evasion technique used by malware. To achieve this, it is crucial to thoroughly investigate the impact on performance as we progressively increase the sequence length of successive API calls under consideration.

Our proposed solution involves mapping each function call directly to a feature in the feature vector, with the value in each position representing the number of times the sample invoked that particular function. This method allows us to create a robust feature representation that is not influenced by the order of API calls, focusing instead on the frequency of each function's invocation.

We structure our experiments into four distinct parts to evaluate this approach comprehensively and ensure the validity of our findings. In the first part, we examine each API call individually, disregarding the context provided by previous and subsequent API calls. We refer to this as the Unigram model, a concept borrowed from natural language processing (NLP). In NLP, a Unigram model analyzes text by considering each word independently, without accounting for the sequence in which words appear. Similarly, in our Unigram model for API calls, we treat each function call as an independent event, counting its occurrences without considering its position in the sequence.

This initial experiment establishes a baseline understanding of how well individual API call frequencies can distinguish between benign and malicious software. By focusing solely on the count of each function, we can determine the effectiveness of this simple yet powerful feature representation in detecting malware. Subsequent parts of our experiments will build upon this foundation, progressively incorporating more contextual information to explore how the performance varies following different lengths of sequences.

By employing this methodical approach, we ensure a comprehensive analysis of the relationship between API call sequences and malware detection accuracy. Our ultimate goal is to identify the optimal balance between feature complexity and detection performance, ultimately developing a robust and efficient malware detection system that is resilient to common evasion tactics.

For the Unigram approach, we simply map each function call to an array index in the feature vector.
We do this by creating a vector, $V$ of length as in~\autoref{eqn:feature_vector}.
\begin{equation}
    |V| = |{Call_1, Call_2, ..., Call_i, ..., Call_n}|
    \label{eqn:feature_vector}
\end{equation}
Each dimension in $V$ corresponds to a specific API Call, and the value of the dimension is the number of API calls belonging to the specific function in a particular sample as shown in~\autoref{api_eqn:call}.
\begin{equation}
    V_i = |Call_i| 
    \label{api_eqn:call}
\end{equation}

When we consider the Bigram model, we also look at the immediately preceding API call for the second part. 
Furthermore, we do this using a sliding window approach over the entire API call sequence.
The length of the feature vector in theory would be the total number of combinations of two API calls, which is $|V|^2$~($59^2$).

Like the Unigram model, we create a feature vector for every combination of API calls.
The vector would comprise two consecutive calls we concatenate, like $Call_1Call_2$.
The total count will be the value for the feature at index $i$.
\begin{equation}
    V_i = |\theta(Call_{i-1},Call_{i})|
\end{equation}
where,
$\theta$ is a mapping from two consecutive API calls to an index.
For the third part, we consider two previous API calls for the sequence. 
%The feature vector is quite large compared to the feature vectors of the unigram and bigram models.
We follow the same procedure as the Unigram model with the only difference in the length of the feature vector and the number of API calls considered.

And for the Trigram model, we consider three consecutive calls as given in~\autoref{eqn:trigram}. In this case, the size of the feature vector expands to $|V|^3$~($=59^3$)~which is quite large compared to the feature vectors of the Unigram and Bigram models.
\begin{equation}
    V_i = |\theta(Call_{i-2},Call_{i-1},Call_{i})|
    \label{eqn:trigram}
\end{equation}
where,
$\theta$ is a mapping from three consecutive API calls to an index.
The index mapping the unique function sequences to an index is provided as a JSON file in our online code repository.
The final model, which we call the Combined model, consists of creating a feature vector that concatenates the feature vectors of the Unigram, Bigram, and Trigram models. 
We do this in the hope that the~\textit{Combined Model}~exploits the positive aspects of Unigram, Bigram, and Trigram models independently to obtain discriminating information from any of the inputs of the models.
%We observe that there are 59 unique API calls to the ntdll.dll library corresponding to the fifty-nine functions we mapped.
%Using these fifty-nine unique function calls, we find that there are 2540 unique~(where we consider two consecutive API calls)~API calls and 5483 unique trigrams ~(where we consider three consecutive API calls)~API calls using a sliding window approach.
The feature vector length for the Unigram model is $59$ corresponding to the number of traced functions to the ntdll.dll, which is manageable, but the Bigram and Trigram models have theoretical lengths of $3481$ and $205379$. 
Although the Bigram feature vector is manageable, it is still quite large, and a Trigram-based feature vector is only possible for more than $330k$ samples on machines with substantial amounts of memory.
Moreover, such a feature vector could be sparse since most values would be zero. 
Therefore, we efficiently identify all the unique bigram and trigram function calls and create feature vectors using only those present in the dataset.
There are~$2540$~unique bigram calls and~$5483$~unique trigram combinations in the dataset.
Therefore, for practicality and to save memory, we limit the Bigram model and Trigram model feature vectors to a length of~$2540$ and $ 5483$~respectively.
We then train a random forest on these feature vectors and predict whether a given sample from the test set is malware or benign.
The dataset is unbalanced, containing many more malicious samples than benign ones.

\section{Results}
\label{sec:results}
The potential for a malware process to run indefinitely in a real-world setting underscores the critical importance of early detection, providing security researchers with a crucial advantage.
Therefore, we run our experiment at different API call counts to determine the best count that gives us the best balance between early prediction and performance.
Our experiment was comprehensive, considering a wide range of lengths: {50, 100, 150, 200, 250, 500, 750, 1000, 2500, 5000, 7500, 10000, 20000, 100000}.
Running the experiment for different maximum API calls allows us to understand an executable's minimum required number of API calls to identify it as malicious or benign software.
We plot this against the F1-Score, the harmonic mean of precision and recall. 
We do not consider accuracy in this case as this is a very unbalanced dataset, and simply predicting the majority class gives us an accuracy value of above~$96\%$.

\begin{figure}
    \centering
    \includegraphics[width=0.45\textwidth]{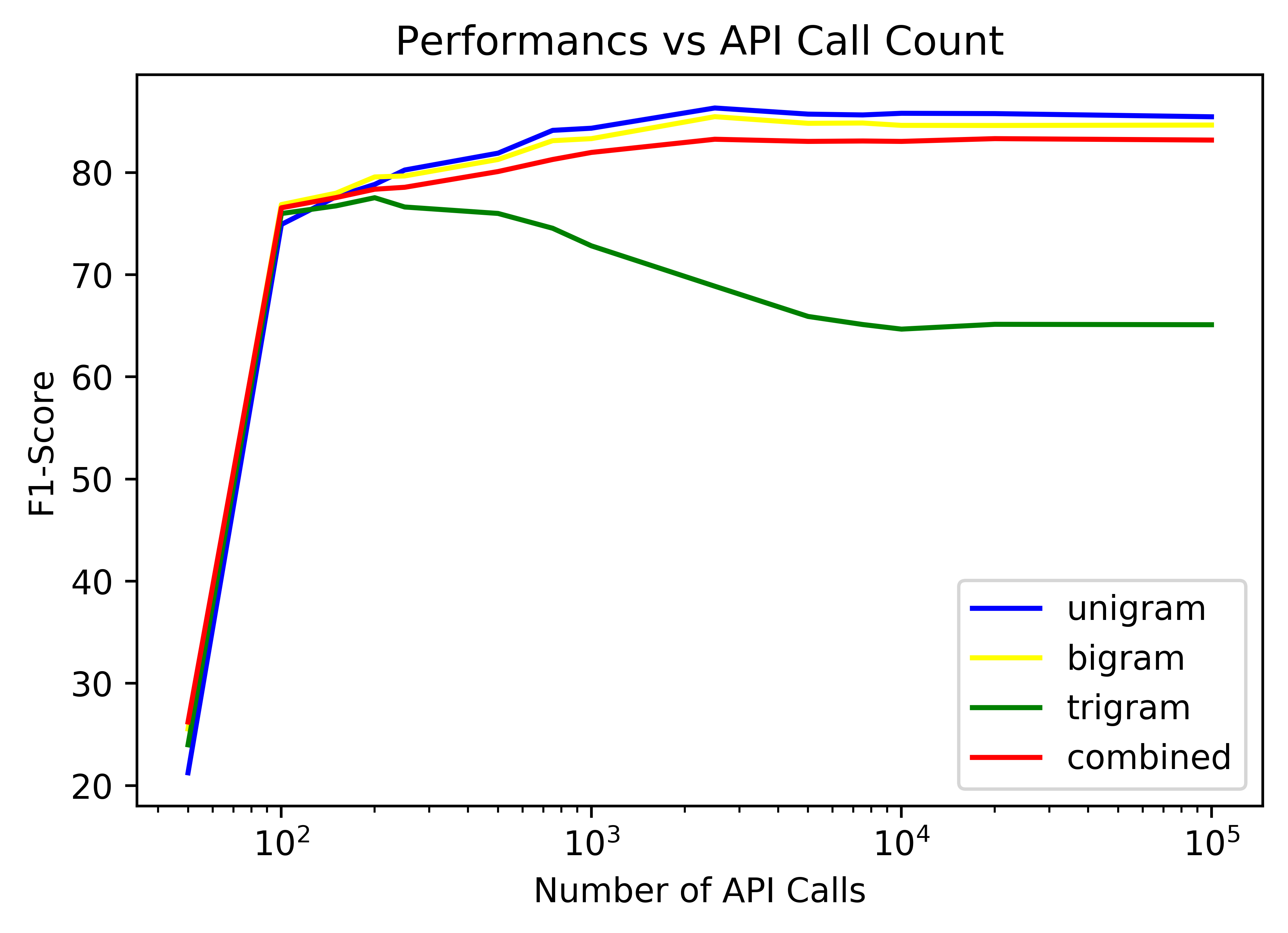}
    \caption{F1-Score for all our models at different max API call counts. The X-axis is on a logarithmic scale. The results are the average of four different runs.}
    %\Description[a short description]{F1-Score for max API calls}
    \label{fig:f1_score}
\end{figure}

\begin{figure}
    \centering
    \includegraphics[width=0.45\textwidth]{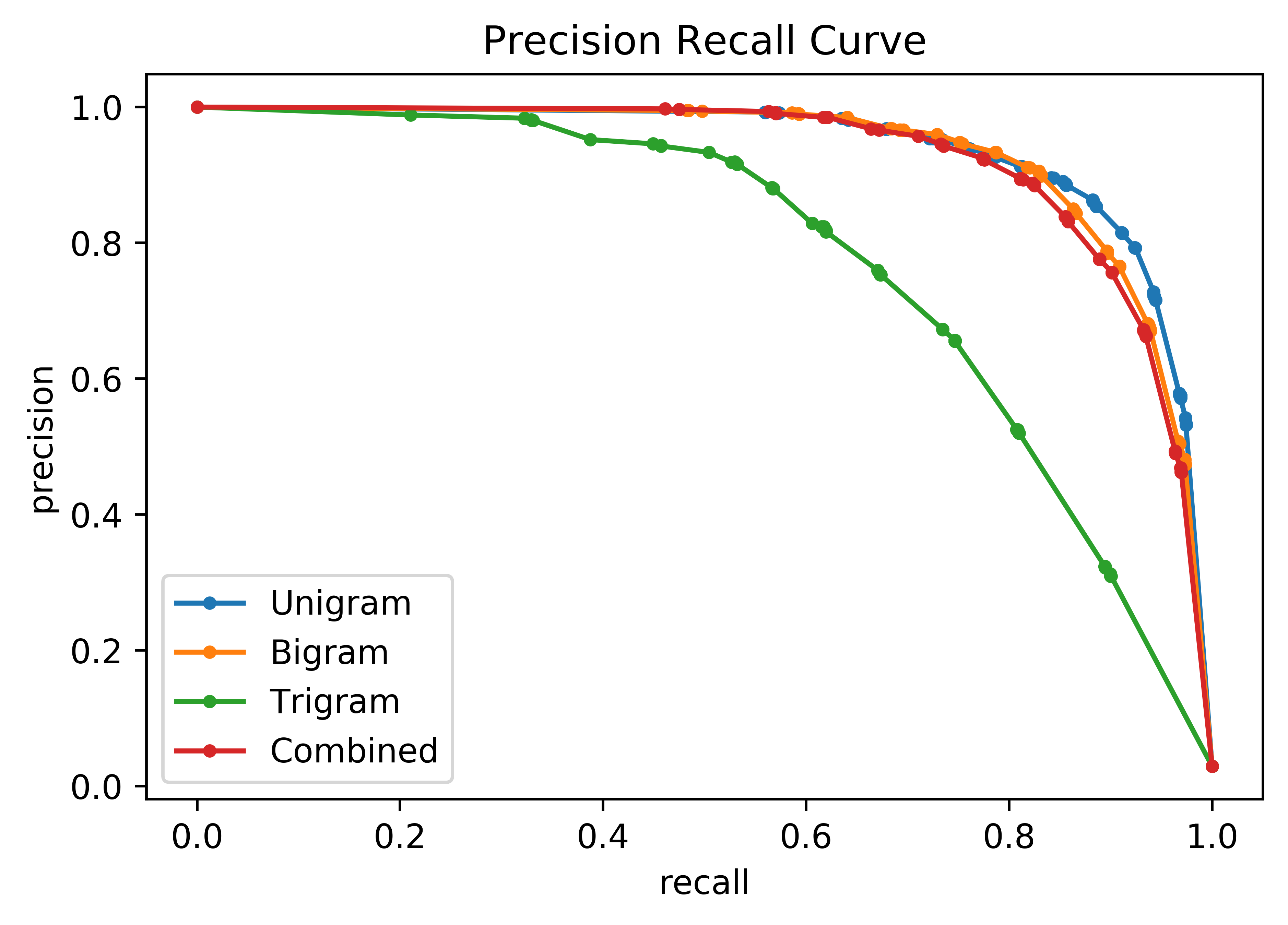}
    \caption{Precision Recall curve for the Unigram, Bigram, Trigram, and combined models. We see that the Trigram model confidence drops very quickly compared to the other models.}
    %\Description[a short description]{PR Curve for the models}
    \label{fig:pr_curve}
\end{figure}

We see from~\autoref{fig:f1_score}~that the metrics are inferior for the initial value of fifty maximum API calls.
Remarkably, the performance sees a substantial leap with just one hundred API calls, highlighting the potential for significant improvement.
It's clear from this analysis that with only a few hundred API calls per sample, we can confidently identify the sample's malicious nature, demonstrating the power of our analysis without the need for temporal information.
Fifty-nine unique API calls to the ntdll.dll library correspond to the fifty-nine functions we mapped.
Using these fifty-nine unique function calls, we find 2540 unique~(where we consider two consecutive API calls)~API calls and 5483 unique trigrams ~(where we consider three consecutive API calls)~API calls using a sliding window approach.
An exciting aspect from~\autoref{fig:f1_score}~is in the trigram-based model~(where we consider a window of three consecutive API calls).
The F1-Score drops off after 200 API calls, and this is due to the large number of unique sequences, which spreads the count out along the feature vector. 
Due to the large number of unique occurrences, the number of calls tends to be unique and the feature vector is a smear of mostly individual unique API calls.
\begin{table}[ht]
    \centering
    \begin{tabular}{|c|c|c|c|c|}
        \hline
        Metric & Unigram & Bigram & Trigram & Combined \\
        \hline
        Accuracy & 99.24 & 99.21 & 98.50 & 98.50 \\
        Precision & 91.04 & 92.47 & 86.71 &  90.91 \\
        Recall & 82.05 & 79.45 & 57.12 & 76.79 \\
        F1-Score & 86.31 & 85.46 & 68.87 & 83.25 \\
        ROC AUC & 0.9843 & 0.9881 & 0.9495 & 0.9812 \\
        \hline
    \end{tabular}
    \caption{Metric values at a maximum of 2500 API calls. We chose 2500 API calls as the maximum limit as it provided the best results across the board.}
    %Unigram refers to only individual API calls, bigram refers to results when considering two consecutive API calls, and trigram refers to the model that considers three consecutive API calls. Combined refers to the model which is trained on the concatenated feature vector input of the unigram, bigram and trigram models. 
    %Our GitHub repository contains the entire set of results.}
    \label{tab:metrics}
\end{table}
In our case, we set the benign class to "1" and the malware class to "0". 
We do this due to highly imbalanced data favoring the malware class. 
In a real-world scenario, we do not want malware samples identified as benign.
Therefore, we need a very precise model so that no malware is identified as benign.
Although having a higher recall is good, in our case, it should not be at the expense of precision.
If our model has false negatives, it means that some benign software was classified as malware and could warrant a closer look.
However, having lower precision might have drastic consequences.
A precision-recall curve can help us understand the model's confidence in test data predictions. 
From~\autoref{fig:pr_curve}, we see that the Unigram model~(only single function call irrelevant of order)~performs very well along with the Bigram, Trigram and combined models.
The only exception is the Trigram model, where the precision drops drastically. 
From this, having a Unigram model should be sufficient for identifying malware from benign software. 
All our results and plots are given in our GitHub repository\footnote{\url{https://github.com/cfellicious/api-based-malware-detection}}.

\iffalse
We also take a look at the~ROC Curve in~\autoref{fig:tpr}~which gives us an idea of the rate of the predictions of true positives to false positives by a model. 
We see that the trigram-based model performs worse than the other models. 
The AUC values in~\autoref{tab:metrics}~also point to the trigram model being the worst.
\begin{figure}[h]
    \centering
    \includegraphics[width=0.45\textwidth]{figures/tpr.png}
    \caption{ROC curve for the unigram, bigram, trigram and combined models. We see that the trigram model performance drops when compared to the other models. AUC values are 0.9843, 0.9881, 0.9495, and 0.9812 for the unigram, bigram, trigram, and combined models respectively.}
    %\Description[a short description]{TPR Curve for the models}
    \label{fig:tpr}
\end{figure}
\fi
From the different metrics and plots, we see that the Bigram model and the Unigram model provide the best performance overall. However, the Unigram model could be favored more simply due to the lower memory requirements and faster execution.

\section{Conclusion}
\label{sec:conclusion}
\iffalse
We created a large corpora of API calls from very recent malware and benign software based on the ntdll Windows library.
Our experiments show that having simply the function call count itself can help us differentiate between malware and benign software to a great degree.
The tests we ran could detect malware with a high degree of certainty.
We also show that for a malware detection system to be effective to a certain degree, we need at least one hundred API calls to the ntdll.dll library.
\fi
We created a substantial corpus of API calls from recent malware and benign software, with a specific focus on the ntdll Windows library, underscoring its significance in our research. 
Our dataset, the largest of its kind available publicly, is a unique and invaluable resource for the research community in malware detection. 
Our experiments demonstrate that even the basic metric of function call counts can significantly distinguish between malicious and benign software. 
The created models show that by using good feature engineering techniques, we can detect malware precisely with negligible performance overhead.
Our method has been proven to detect malware with a small number of API calls, demonstrating its efficiency and practicality. 
Furthermore, we found that for a malware detection system to be effective, it must analyze at least two hundred and fifty API calls to the ntdll.dll library. 
This threshold ensures a reliable level of detection accuracy, reinforcing the importance of comprehensive data collection in developing robust malware detection systems. 
Our findings also highlight the potential of simple yet effective features in developing efficient and scalable malware detection solutions, paving the way for future research and advancements in cybersecurity.
\bibliographystyle{bibliography/splncs04}
\bibliography{bibliography/bibliography}
\end{document}